\documentclass[aps,prd,,groupedaddress,showpacs,showkeys,nofootinbib,amssymb]{revtex4}
\usepackage[T1]{fontenc}
\usepackage[latin1]{inputenc}
\usepackage{graphicx}
\usepackage[english]{babel}
\usepackage{amsmath}
\usepackage{amssymb}
\usepackage{amsfonts}

\begin{document}
\def\pp{{\, \mid \hskip -1.5mm =}}
\def\cL{{\cal L}} 
\def\be{\begin{equation}}
\def\ee{\end{equation}}
\def\bea{\begin{eqnarray}}
\def\eea{\end{eqnarray}}
\def\beq{\begin{eqnarray}}
\def\eeq{\end{eqnarray}}
\def\tr{{\rm tr}\, }
\def\nn{\nonumber \\}
\def\e{{\rm e}}

\title{\textbf{Neutrino oscillation phase dynamically induced by $f(R)$-gravity }}

\author{Salvatore Capozziello, Mariafelicia  De Laurentis, Daniele Vernieri}

\affiliation{\it Dipartimento di Scienze Fisiche, Università
di Napoli {}``Federico II'', INFN Sez. di Napoli, Compl. Univ. di
Monte S. Angelo, Edificio G, Via Cinthia, I-80126, Napoli, Italy}

\date{\today}

\begin{abstract}
The gravitational phase shift of neutrino oscillation can be discussed in the framework of $f(R)$-gravity. We show that the shift of quantum mechanical phase can depend on the given $f(R)$-theory that we choose. This fact is general and  could constitute a fundamental test to discriminate among the various alternative relativistic theories of gravity. Estimations of ratio between the gravitational phase shift and the standard phase are carried out for the electronic Solar neutrinos.

 \end{abstract}
\maketitle

Neutrinos are elementary particles that travel at the speed of light (or close to it if massive), are electrically neutral and are capable of passing through ordinary matter with minimal interaction. Due to these properties, they can be investigated, in principle, 
  at all length scales, ranging from nuclei \cite{Reines}, to
molecular structures \cite{Collar}, up to galaxies \cite{Weinberg,CapLam} and to the whole Universe \cite{Blasone}. 
They results from  radioactive decays or nuclear reactions such as those that take place in the Sun, stars or
 nuclear reactors. In particular, they are generated   when cosmic rays hit atoms.
Current evidences of dark matter and dark energy can be related to the issue that
neutrinos have masses and that  mass eigenstates mix and/or superimpose \cite{Barenboim, Barenboim2, Barenboim3, nieuwen}.
The observation of such a mixing is related to suitable  constraints.  Such constraints should work  on
 observables sensitive to  the effective neutrino
mass as the mass in Tritium-beta decay, the sum of neutrinos masses in cosmology and  the
effective Majorana neutrino mass in neutrinoless double-beta decay  \cite{Fogli}. 

A key role  is played by the neutrino oscillations
that allow the transition among the three  types or {\it "flavor"} eigenstates, that is the electron, muon, and tauon neutrinos.
It is well known that such a problem is still open and the research of new effects,
in which the oscillations could manifest is one of the main goal of modern
physics. For this reason, the quantum mechanical phase of neutrinos, propagating
in gravitational field, has been discussed by several authors, also in view of the
astrophysical consequences. More controversial is the debate concerning the redshift
of flavor oscillation clocks, in the framework of the weak gravitational field
of a star \cite{Ahluwalia}. It has also been suggested that the gravitational oscillation phase
might have a significant effect in supernova explosions due to the extremely large
fluxes of neutrinos produced with different energies, corresponding to the flavor
states. This result has been confirmed in \cite{Grossman}, and it has been also derived under
the assumption that the radial momentum of neutrinos is constant along the trajectory
of the neutrino itself \cite{Konno}. Besides, neutrino oscillations, in particular the
gravitational part of the oscillation phase, could straightforwardly come into the
debate, which is recently risen, to select what is the correct theory of gravity, due
to the well known experimental and theoretical shortcomings of General Relativity
\cite{Will,gaetano}. 

On the other hand, higher-order theories, extending in some way General Relativity, 
allow to pursue different approaches \cite{mauro,zerbini,odintsov,farasot}. 
This viewpoint does not require to find out
candidates for dark energy and dark matter at fundamental level
(they have not been detected up to now) but it takes into account
only the ``observed'' ingredients ({\it i.e.} gravity, radiation, neutrinos and
baryonic matter). However,  the {\it l.h.s.} of the Einstein equations has to
be extended and modified. Despite of this modification, it is in
agreement with the spirit of General Relativity since the only
request is that the Hilbert-Einstein action should be generalized
asking for a gravitational interaction  acting, in principle,  in
different ways at different scales \cite{mnras1,mnras2}.  This feature could be extremely interestindg for neutrino oscillations since further gravitational interaction lengths, emerging from extended theories of gravity, could be related to neutrino oscillation phase. On the other hand, the experimental identification of such a gravitational phase could be a formidable probe both for confirming or ruling out such theories at a fundamental level.

In this letter, after a short review of the gravitational phase shift in neutrino oscillations, we briefly outline the theory of $f(R)$-gravity putting in evidence the Yukawa-like correction to the gravitational potential emerging, in general, as soon as $f(R)\neq R$, that is the theory is not General Relativity. Finally, we discuss the quantum mechanical oscillation phase shift for propagation of neutrino in a generic analytic $f(R)$-gravity model.

\vspace{3mm}

Let us start our discussion considering  how the gravitational field contributes to the neutrino oscillations. The approach has been firstly developed in \cite{Ahluwalia} and we will outline the main results reported there.
If $R_A$ is the size of a physical region where neutrinos are generated, a neutrino energy eigenstate $E_{\nu}$ can be denoted by $|\nu_l,R_A\rangle$ (where l = $e,\mu,\tau$ represents the weak flavor eigenstates). The three neutrino mass eigenstates can be represented by $|\nu_i \rangle$ with $i=1,2,3$ corresponding to the masses $m_1, m_2, m_3$. The mixing between mass and flavor
eigenstates is achieved by the unitary transformation
\begin{equation}
|\nu_l',R_A\rangle=\sum_{i=1,2,3}U_{li}|\nu_i\rangle\label{15},
\end{equation}
where
\begin{equation}
U\left(\theta,\beta,\psi\right)=\left(\begin{array}{ccc}
c_{\theta}c_{\beta} & s_{\theta}c_{\beta} & s_{\beta}\\
-c_{\theta}s_{\beta}s_{\psi}-s_{\theta}c_{\psi} & c_{\theta}c_{\psi}-s_{\theta}s_{\beta}s_{\psi} & c_{\beta}s_{\psi}\\
-c_{\theta}s_{\beta}c_{\psi}-s_{\theta}s_{\psi} & -s_{\theta}s_{\beta}c_{\psi}-c_{\theta}s_{\psi} & c_{\beta}c_{\psi}\end{array}\right)\label{16}
\end{equation}
is a $3\times3$ unitary matrix parametrized by the three mixing angles $\eta =\theta, \beta,\psi$ with
$c_\eta=\cos\eta$ and $s_\eta=\sin\eta$. At time $t = t_B > t_A$, the weak flavor eigenstates
can be detected in a region $R_B$ and, in general, the evolution is given by
\begin{equation}
|\nu_{l},R_{B}\rangle=exp\left(-\frac{i}{\hbar}\int_{t_{A}}^{t_{B}}\mathcal{H}dt+\frac{i}{\hbar}\int_{r_{A}}^{r_{B}}\overrightarrow{P}\cdot d\overrightarrow{x}\right)|\nu_l,R_A\rangle\label{17},
\end{equation}
where $\mathcal{H}$ is the Hamiltonian operator associated to the system  representing the time translation
operator and $\overrightarrow{P}$ is the momentum operator  representing the spatial translation operator.
The phase change in Eq.(\ref{17}) is  the argument of the exponential function. It can be
recast in the form
\begin{equation}
\phi_{\nu}=\frac{1}{\hbar}\int_{r_{A}}^{r_{B}}\left[E\frac{dt}{dr}-p_{r}\right]dr\label{18}\,.
\end{equation}
 The covariant formulation is
\begin{equation}
\phi_{\nu}=\frac{1}{\hbar}\int_{r_A}^{r_B}mds=\frac{1}{\hbar}\int_{r_A}^{r_B}p_{\mu}dx^{\mu},\label{19}
\end{equation}
where $p_\mu=mg_{\mu\nu}\frac{dx^\nu}{ds}$ is the $4$-momentum of the particle. The effect of
 gravitational field is given by $g_{\mu\nu}$ and, in general, the neutrino oscillation
probability from a state $|\nu_l,R_A\rangle$  to another state $|\nu_l,R_B\rangle$	 is given by
\begin{eqnarray*}
\mathcal{P}\left[\left|\nu_{l},R_{A}\right\rangle \rightarrow\left|\nu_{l'},R_{B}\right\rangle \right] & = & \delta_{ll'}-4U_{l'1}U_{l1}U_{l'2}U_{l2}\sin^{2}\left[\phi_{0}^{21}+\right.\\ 
 &  &+\left.\phi_{G}^{21}\right] -4U_{l'1}U_{l1}U_{l'3}U_{l3}\cdot\\
 &  &\cdot \sin^{2}\left[\phi_{0}^{31}+\phi_{G}^{31}\right] -4U_{l'1}U_{l1}U_{l'3}\cdot \\
 &  &\cdot U_{l3}\sin^{2}\left[\phi_{0}^{31}+\phi_{G}^{31}\right]\label{20},\end{eqnarray*}
 
 where $\phi_{0}^{ij}$ are the usual kinematic phase while $\phi_{G}^{ij}$ are the gravitational contributions. It can be shown that, in a flat space-time, the $\phi_{G}^{ij}$ contributions are zero. In fact, a particle passing nearby a point mass feels a Schwarzschild geometry so  the trajectories is
\begin{equation}
dx\simeq\left[1-\frac{2G_{N}M}{c^{2}r}\right]cdt\label{21}\,.\end{equation}
If the effects of gravitational field are vanishing, Eq. (\ref{21})  becomes $dx\simeq cdt$. Considering two generic neutrino mass eingestates in a Schwarzshild geometry, the standard phase of neutrino oscillation is

\begin{equation}
\phi_0=\frac{\Delta m^2 c^3}{4E \hbar}\left(r_B-r_A\right),       \label{23}
\end{equation}
while the total gravitational phase shift is

\begin{equation}
\phi_{grav}=\frac{G_{N}\bigtriangleup m^{2}M c}{4\hbar E}\log\frac{r_{B}}{r_{A}}\,,\label{22}\end{equation}
as shown in \cite{Ahluwalia}, where $\bigtriangleup m^{2}$ is the mass squared difference, $\bigtriangleup m^{2}=\left|m_{2}^{2}-m_{1}^{2}\right|$, $E$ the neutrino energy, $r_A$ and $r_B$ the point where neutrinos are created and detected, respectively. Nevertheless, assuming that the neutrino energy is constant along the trajectory,  the term (\ref{22}) could be cancelled out at  typical astrophysical scales \cite{Bhattacharya}.

With this considerations in mind, let us take into account how possible corrections to the Newtonian potential could affect this result. 
We will  follow the discussion developed in \cite{noi-prd,noimpla}.

Let us consider the general gravitational action

\begin{equation}\label{actfR}
{\cal A}\, = \,\int
d^4x\sqrt{-g}\left[f(R)+{\cal X}{\cal L}_m\right]\,,
\end{equation}
where $f(R)$ is an  analytic function of the Ricci scalar $R$, $g$ is the
determinant of the metric $g_{\mu\nu}$, ${\displaystyle {\cal X}=\frac{16\pi
G_{N}}{c^4}}$ is the coupling constant and ${\cal L}_m$ is
the  perfect-fluid matter Lagrangian. Such an
action is the straightforward generalization of the
Hilbert-Einstein action of General Relativity obtained for $f(R)=R$.
In  the  metric approach \cite{mauro}, the field equations
are obtained by varying (\ref{actfR}) with respect to the
metric\,:

\begin{eqnarray}\label{fe}
f'(R)R_{\mu\nu}-\frac{1}{2}f(R)g_{\mu\nu}-f'(R)_{;\mu\nu}+g_{\mu\nu}\Box
f'(R)=\frac{\mathcal{X}}{2}T_{\mu\nu}\,,
\end{eqnarray}
that are fourth-order field equations in the metric derivatives.
 $T_{\mu\nu}=\frac{-2}{\sqrt{-g}}\frac{\delta(\sqrt{-g}{\cal
L}_m)}{\delta g^{\mu\nu}}$ is the energy momentum tensor of
matter, the prime indicates the derivative with respect to $R$ and
$\Box={{}_{;\sigma}}^{;\sigma}$ is the d'Alembert operator. We adopt the signature
$(+,-,-,-)$.
We do not want to  impose a particular forms for  $f(R)$-model but
only consider analytic Taylor expansion where the cosmological term and terms higher than second are discarded. The Lagrangian is then
\begin{equation}
\label{f}
f(R) \sim a_1 R + a_2 R^2 + ... 
\end{equation}
where the parameters $a_{1,2}$ specifies the particular models. Performing the Newtonian limit of the above field equations with the theory given by Eq.(\ref{f}), it is possible to demonstrate, in general, that 
 the gravitational potential, generated
by a point-like matter distribution is (see \cite{noi-prd,noimpla} for details):
\begin{equation}
\label{gravpot1} \Phi(r) = -\frac{3 G_{N} M}{4 a_1
r}\left(1+\frac{1}{3}e^{-\frac{r}{L}}\right)=\Phi(r)_{Newton}+\Phi(r)_{Yukawa}\,,
\end{equation}
where
\begin{equation}\label{lengths model}
L \equiv L(a_{1},a_{2}) =  \left( -\frac{6 a_2}{a_1}
\right)^{1/2}\,.
\end{equation}
$L$ is an  {\it interaction gravitational length}  due to the correction to the Newtonian potential. However, as soon as $a_1= 3/4,$  $a_2 =0$ and $\Phi(r)_{Yukawa}\rightarrow 0$,  the standard Newtonian limit of General Relativity is fully recovered (for a discussion on this point, see \cite{mnras2}).

Now we calculate the gravitational phase shift of neutrino oscillation in $f(R)$-gravity using the potential (\ref{gravpot1}) in the Eq. (\ref{18}). We obtain the general expression

\begin{equation}
\phi_{grav}=\frac{\Delta m^2 M c}{4 \hbar E}\left( \frac{3 G_{N}}{4 a_{1}}\right)\int_{r_A}^{r_B}\left(\frac{1}{r}+\frac{1}{3r} e^{-\frac{L}{r}}\right),
\end{equation}

from which we have the following result
\begin{equation}
\phi_{grav}=\frac{\Delta m^2 M c}{4 \hbar E}\left( \frac{3 G_{N}}{4 a_{1}}\right)\left[ \log \frac{r_B}{r_A}+ \sum_{n=1}^{\infty}(-1)^{n}\left(\frac{\left(\frac{r_B-r_A}{L}\right)^n}{n \cdot n!}\right)\right]
=\phi_{Newton}+\phi_{Yukawa}\, ,
\end{equation}

where $\phi_{grav}$ is the total gravitational phase shift in Eq. (\ref{22}) and 
\begin{equation}
\label{corr}
\phi_{Yukawa}=\frac{\Delta m^2 M c}{4 \hbar E}\left( \frac{3 G_{N}}{4 a_{1}}\right)\left[ \sum_{n=1}^{\infty}(-1)^{n}\left(\frac{\left(\frac{r_B-r_A}{L}\right)^n}{n \cdot n!}\right)\right]\,  .
\end{equation}
The Yukawa term disappears in standard Einstein gravity, that is $f(R)=R$.
Note that the series in above equation is absolutely convergent. If we consider Solar neutrinos,  we can  use the following values: $M\sim M_\odot \sim 1.9891\times 10^{30}$Kg, $r_{A}\sim r_{\oplus}\sim6.3 \times 10^{3}$Km, and $r_{B}\sim r_A+D$, where $D\sim 1.5\times 10^8$Km is the Sun-Earth distance. 
In order to estimate the  phases differences (\ref{23}), (\ref{22}) and (\ref{corr}),  we introduce the ratio $Q_{grav}$ defined as
\begin{equation}
Q_{grav}=\frac{\phi_{Newton}}{\phi_0}\sim \frac{G_N M \log \frac{r_B}{r_A}}{c^2 (r_B-r_A)}\sim 10^{-7},   
\end{equation}
and the ratio $Q_{Yukawa}$ defined as
\begin{equation}
Q_{Yukawa}=\frac{\phi_{Yukawa}}{\phi_0}\sim \frac{G_N M}{c^2 (r_B-r_A)}\sum_{n=1}^{\infty}(-1)^{n}\left(\frac{\left(\frac{r_B-r_A}{L}\right)^n}{n \cdot n!}\right), \label{fi}
\end{equation}
where we have assumed that $3/4a_1\sim1$. Note that both $Q_{Newton}$ and  $Q_{Yukawa}$ do not depend on the squared-mass difference $\Delta m^2$ and on the neutrino energy $E$. The ratio $Q_{Yukawa}$ can be calculated for different values of the {\it interac-\\tion lenght} $L$.  
For example from Eq. (\ref{fi}), after summing the series, we obtain the results:
\begin{equation}
L\sim 1.5\cdot10^7 Km \Longrightarrow Q_{Yukawa}\sim -2.9\cdot10^{-8},
\end{equation}
\begin{equation}
L\sim 1.5\cdot10^8 Km \Longrightarrow Q_{Yukawa}\sim -8\cdot10^{-9}.
\end{equation}
In this way the values of $Q_{Yukawa}$  can be seen as corrections to the standard gravitational phase shift of neutrino oscillations depending on the particular choice of $L$ and so, through Eq.(\ref{lengths model}), directly on the particular $f(R)$-model considered.

We remark that the calculated correction to gravitational phase shift in Eq. (\ref{corr}) depends on the  {\it interaction lenght} $L$ defined in Eq. (\ref{lengths model}).  This is directly related to the particular  $f(R)$-gravity model through the coefficients $a_1$ and $a_2$ in Eq. (\ref{f}). This fact could be used as an experimental test in order to probe a  given gravity theory through the neutrino oscillation induced by means of the gravitational field itself. On the other hand,   interpreting $L$ as the characteristic wavelenght of the neutrino interaction with the gravitational field, the gravitational phase correction could be used as a method to constrain the mass of electronic neutrinos travelling from the Sun to the Earth surface, or, eventually,  also from other neutrinos sources as Supernovae or neutron stars. 

%%%%%%%%%%%%%%%%%%%%%%%%%%%%%%%%%%%%%%%%%%%%%%%%%%%%%%%%%

\end{document}